\documentclass[manuscript,screen]{acmart}
\AtBeginDocument{%
  \providecommand\BibTeX{{%
    \normalfont B\kern-0.5em{\scshape i\kern-0.25em b}\kern-0.8em\TeX}}}

\usepackage{wasysym}
\usepackage{tabularx}

\newcommand{\gender}{\textsc{Gender}}
\newcommand{\age}{\textsc{Age}}

\setcopyright{rightsretained}
\copyrightyear{2023}
\acmYear{2023}

\acmConference[]{Manuscript submitted to ACM}{2023}{}
\acmBooktitle{Manuscript submitted to ACM} 
\acmPrice{}
\acmISBN{}

\begin{document}

\title{Gender, Age, and Technology Education Influence the Adoption and Appropriation of LLMs}

\author{Fiona Draxler}
\email{fiona.draxler@ifi.lmu.de}
\orcid{0000-0002-3112-6015}
\affiliation{%
  \institution{LMU Munich}
  \city{Munich}
  \country{Germany}
  \postcode{80337}
}

\author{Daniel Buschek}
\orcid{0000-0002-0013-715X}
\email{Daniel.Buschek@uni-bayreuth.de}
\affiliation{%
  \institution{University of Bayreuth}
  \city{Bayreuth}
  \country{Germany}}

\author{Mikke Tavast}
\email{mikke.tavast@aalto.fi}
\orcid{0000-0003-0671-5755}
\affiliation{%
  \institution{Aalto University}
  \city{Espoo}
  \country{Finland}
}

\author{Perttu Hämäläinen}
\email{perttu.hamalainen@aalto.fi}
\orcid{0000-0001-7764-3459}
\affiliation{%
  \institution{Aalto University}
  \city{Espoo}
  \country{Finland}
}

\author{Albrecht Schmidt}
\email{albrecht.schmidt@ifi.lmu.de}
\orcid{0000-0003-3890-1990}
\affiliation{%
  \institution{LMU Munich}
  \city{Munich}
  \country{Germany}
  \postcode{80337}
}

\author{Juhi Kulshrestha}
\email{juhi.kulshrestha@aalto.fi}
\orcid{0000-0002-4375-4641}
\affiliation{%
  \institution{Aalto University}
  \city{Espoo}
  \country{Finland}
}

\author{Robin Welsch}
\email{robin.welsch@aalto.fi}
\orcid{0000-0002-7255-7890}
\affiliation{%
  \institution{Aalto University}
  \city{Espoo}
  \country{Finland}
}

\renewcommand{\shortauthors}{Draxler et al.}

\def\subsectionautorefname{Section}
\def\subsubsectionautorefname{Section}
\def\sectionautorefname{Section}

\begin{abstract}
Large Language Models (LLMs) such as ChatGPT have become increasingly integrated into critical activities of daily life, raising concerns about equitable access and utilization across diverse demographics. This study investigates the usage of LLMs among 1,500 representative US citizens. Remarkably, 42\% of participants reported utilizing an LLM. Our findings reveal a gender gap in LLM technology adoption (more male users than female users) with complex interaction patterns regarding age.
Technology-related education eliminates the gender gap in our sample.
Moreover, expert users are more likely than novices to list professional tasks as typical application scenarios, suggesting discrepancies in effective usage at the workplace.
These results underscore the importance of providing education in artificial intelligence in our technology-driven society to promote equitable access to and benefits from LLMs. We urge for both international replication beyond the US and longitudinal observation of adoption.  
\end{abstract}



\maketitle

\section{Introduction}
A significant recent development in Artificial Intelligence (AI) is the advent of Large Language Models (LLMs) such as ChatGPT and GPT-4. These LLMs support users in various tasks, including information search, coding, and creative writing \cite{brown2020language,openai2023gpt4,unesco2023ChatGPT}. Chat-based interfaces provide seemingly simple access for novice and expert users alike.
However, the literature is relatively scant regarding systematic reports on the profiles of non-professional users of LLMs, typical tasks and usage scenarios, and adoption barriers. Nonetheless, understanding demographic patterns in LLM usage is an important prerequisite for enabling equitable access to LLMs and leveling imbalances in an increasingly AI-reliant society.

Studies on technology adaptation and acceptance typically find that age and gender predict technology usage \cite{morris2000age,robinson2015,venkatesh_why_2000}. Thus, technological transformation %
does not penetrate society equally but is accelerated in some demographic profiles. This amplifies societal challenges such as the gender pay gap or the employability crisis in old adults \cite{robinson2015}.
Considering the rapid development and proliferation of LLMs, it is essential to evaluate and address demographic challenges as early as possible.

To do so, we compare the demographics of users and non-users of LLMs in a US sample representative of the population in terms of age, gender, and ethnicity ($n = 1,500$). Moreover, we explore  typical LLM scenarios for novices and experts and barriers to LLM adoption.
We hypothesized that gender and age differences will predict the use of LLMs.

Our study (run four months after the release of ChatGPT)  shows that females are less likely to use LLMs than males and that younger age groups are more likely to use LLMs than older ones. There is a complex interaction pattern of age and gender. The gender gap is most pronounced in young adults and is only marginal in old age. This is mirrored in the frequency of use, where most young females use LLMs once a month or less often, while male users of LLMs report a higher frequency of use.
We also show that the gender effect is diminished for populations with degrees in technology, even if accounting for differences in levels of education.
Typical usage scenarios suggest that experts are more likely to use LLMs for professional tasks, such as generating formal texts and coding, while novices try out LLM capabilities and use it for entertainment purposes.
Finally, non-users reported a variety of reasons for not using LLMs including a lack of knowledge and needs, ethical concerns, and the impersonal interaction.

Based on these findings, we contribute recommendations to promote LLM adoption in disadvantaged demographic profiles.
Notably, in line with organizations such as IDB, OECD, and UNESCO \cite{AIWorkingLivesofWomen, west2019d}, we call for education in technological disciplines with a focus on AI early on. %
Moreover, ethical concerns show a need for differentiated and contextualized discussions beyond all-or-nothing conclusions, e.g., when considering LLMs for educational usage.
Our study also sets the basis for the advancement of demographic studies on technological adaptation beyond the US and in a longitudinal fashion.

\section{Related Work}

We relate our work to studies of technology adoption that indicate demographic effects, %
and also consider LLM-specific factors such as characteristics of common application domains and the interaction with chat-based interfaces.

\subsection{Demographic Factors and the Adoption of Novel Technologies}
\label{sec:rw_demographics}

Demographic studies show that age and gender influence technology adoption.
Younger individuals readily adopt novel technologies, often driven by their ability to learn interaction patterns and their openness to experiment \cite{parasuraman_updated_2015}.
Conversely, the perceived ease or difficulty of a given system predicts adoption in old age \cite{morris2000age}. %
For example, younger generations use mobile devices more frequently and perceive them as less complex and more visible than older generations \cite{magsamenconrad_mobile_2020}.
Furthermore, there is a gender gap in technology adoption, with males generally engaging with new technologies earlier than females \cite{west2019d}. This may be attributed to factors such as differences in self-efficacy and education in technological disciplines. Women have also been found to be more strongly influenced by social norms than men \cite{venkatesh_why_2000}. %
For example, previous studies found a higher acceptance of AI in banking services \cite{tubadji_cultural_2021} and higher levels of immersion in VR \cite{felnhofer2012virtual} for male than female participants.
Additional factors such as socio-economic status and cultural background also predict technology adoption. These are not evaluated in this paper but for an overview see, for example \cite{tubadji_cultural_2021,rojasmendez_demographics_2017,van_deursen_digital_2021}.

The demographics of LLM users may exhibit more complex patterns than is the case for other technologies:
Other studies have focused on larger timeframes of years, often within professional contexts \cite{morris2000age}, and may not describe disruptive technologies. LLMs have disrupted society within months, which may influence related demographic patterns.

Knowing the impact of demographic factors is essential for achieving equitable use of novel technologies by building up expertise and lowering access barriers. \citet{magsamenconrad_mobile_2020} report that inter-generational communication helps overcome age barriers, while \citet{west2019d} postulate technology education for closing the gender gap.
Moreover, efforts under the umbrella of the United Nations' sustainability development goals 4 (Quality Education) and 5 (Gender Equality) further push equity.

\subsection{Application Domains of LLMs} %

LLMs can answer questions \cite{brown2020language}, solve coding tasks \cite{zan-etal-2023-large,chen2021humaneval}, assist learning \cite{kasneci_chatgpt_2023}, perform creative writing assignments \cite{lee_coauthor_2022}, and more.
These diverse abilities emerge from one main training objective: Predict the next token---a word piece---in a large and diverse text corpus  \cite{brown2020language,wei2022emergent}, although performance can be further improved by reinforcement learning from human feedback and/or finetuning using curated examples of desired human-AI interactions  \cite{ouyang2022training,zhou2023lima}. %
The rapid development has also sparked innovation in professional contexts. For instance, business professionals harness LLMs to automate customer service and provide personalized customer recommendations \cite{paul_span_2023}. Likewise, researchers use LLMs to analyze and synthesize the verbal data of participants \cite{hamalainen2023evaluating} but also to generate proofs to theorems \cite{han2022proof} and design and analyze proteins \cite{madani2023large,lin2023evolutionary}. %
Initial research suggests that integrating LLMs into work processes can significantly increase efficiency \cite{eloundou_gpts_2023}. Conversely, this means that people who do not use LLMs, or are inefficient in doing so, will not be able to compete in the long run, resulting in an economic disadvantage.

Both professionals and non-professionals can access LLMs online \cite{openai2023gpt4}.
In this context, our analysis of typical usage scenarios and differences between novice and expert users sheds light on contexts that need to be prioritized for supporting effective use across domains and experience levels.

\subsection{Interaction with LLMs}

Thanks to simple chat interfaces in web browsers and mobile apps, LLMs enable users to leverage AI using natural language without having to learn or adopt new interaction modalities \cite{folstad2017chatbots}. Thus, LLMs have the potential to democratize technology access for daily tasks.
Despite this, novice users still struggle with formulating prompts \cite{zamfirescupereira_why_2023}. The demand for effective prompting also shows in a new market for prompt engineers who specialize in crafting prompts for interacting with LLMs \cite{popli_ai_2023}.

In sum, current interaction modalities promise easy initial access for a large share of the population. However, the lack of guidance could easily frustrate users with little knowledge of prompting, restricting effective use beyond simple tasks.
Here, our analyses of adoption barriers and typical scenarios for novices and proficient users can inform the design of more widely accessible interfaces, e.g., specialized interfaces that provide more guidance for a given task.

\section{Method}

We conducted an online survey to assess socio-demographic characteristics related to LLM usage. We also collect typical usage scenarios and reasons for not using LLMs. The procedure was reviewed and approved by the local ethics committee. %
The measures and data set are available on OSF: \url{https://osf.io/bdn9p/?view_only=1a44dfd53cc1442a87bfc3f49560b112}.

\subsection{Hypotheses and Pre-Registration}
\label{sec:hypotheses}

Following related work on adoption and estimated usage statistics of novel technologies (cf. \autoref{sec:rw_demographics}), we predict the following relationships between gender, age, and LLM usage (pre-registered at \url{https://aspredicted.org/VCN_CCS}):
\begin{enumerate}
    \item[H1:] Men are more likely to use LLMs than people of other genders.
    \item[H2:] Younger age groups are more likely to use LLMs than older age groups.
    \item[H3:] There is an interaction effect of gender and age for the likelihood of using LLMs.
\end{enumerate}

For all three, we analyze binary usage (\textit{yes}/\textit{no}) followed by usage frequency for participants that reported LLM use (i.e., \textit{yes}).
Subsequently, we perform an exploratory analysis of self-declared expertise and technology education as possible mediating factors.
Moreover, we explore adoption barriers and typical scenarios for novice and competent users. %

\subsection{Participants}
\label{sec:participants}
We recruited a nationally representative US sample of 1,500 adults aged 18+ years in Prolific (maximum sample size). One revoked consent, two were bots, and two failed an attention check. Excluding these, we analyzed data from 1,495 participants.
They declared their gender as male (721), female (760), non-binary/diverse (11), self-described as agender (1), and two preferred not to disclose. We only analyzed participants declaring to be male or female because the numbers of other genders were too low for reliable numeric assessments; however, the OSF repository includes a descriptive summary of LLM usage. Ethnicity, according to the simple ethnic groups defined by the UK Office of National Statistics \cite{ethnicity_uk_2022}, was White (1167), Black (190), Asian (87), Mixed (29), and Other (22), also see OSF for a summary.
They self-reported their education level as ``Some high school or less'' (6), ``High school diploma or GED'' (181), ``Some college but no degree'' (273), ``Associate or technical degree'' (192), ``Bachelor's degree'' (557), ``Graduate or professional degree'' (284). Two did not disclose their education level.
Age ranged from 18 to 93 years ($ M = 45.80 $, $ SD = 15.79 $).

\subsection{Procedure}
\label{sec:procedure}
The survey queried socio-demographics, including educational background, occupation, household income, and political stance. Following a brief explanation of LLMs and a list of examples\footnote{The list of examples was: ``GPT-3, ChatGPT, GPT-4, Bing with GPT integration, Bard, BERT, and tools such as LAIKA and Typing Mind''}, participants then reported if they had ever used an LLM. We did not differentiate by types of LLMs.
If they had, we asked for their usage frequency, self-assessed experience, and usage scenarios. If not, we asked for envisioned usage scenarios and their reasons not to use LLMs.  %
A full list of measures is provided on OSF.
The median completion time was 4:15 minutes.

\subsection{Data analysis}
\label{sec:data_analysis}

We apply Bayesian regression, inductive coding, and topic modeling:
For modeling LLM adoption, we used Bayesian linear mixed models and a logistic link function. We estimated odds (ratios) and quantify uncertainty based on the information in our data and the priors applied. We used brms~\citep{burkner2017brms} for modeling, a wrapper for the STAN sampler \citep{carpenter_stan_2017}. %
We computed 4 Hamilton-Monte-Carlo chains with 40000 iterations each and 10\% warm-up samples. %
We calculated $p_b$ as the relative proportion of posterior samples being zero or opposite to the median (meaningful effect: $p_b \leq 2.5\%$). %
This quantifies the proportion of probability that the effect is zero or opposite given the data observed. %
We calculated the median and the Highest-Density Interval (HDI) at 95\% of the posterior distribution for all parameters, which indicates the possible range of effects given the data. We run posterior predictive checks to analyze interaction effects.

For categorizing usage scenarios, we inductively coded the open-ended responses. Rater 1 defined an initial categorization scheme by screening the data. Rater 2 refined the scheme and categorized all 626 scenarios (multiple assignments possible). Raters 3 and 4 provided a second categorization for 461 and 165 responses, respectively.
The interrater agreement was 83.9\% for Raters 2 and 3 and 83.7\% for Raters 2 and 4\footnote{Cohen's kappa was not used because of the large skew in the data: negative assignments to a class are much more likely than positive assignments \cite{cicchetti_high_1990}}. %
For the final assignment, Rater 1 independently re-coded all class assignments where R2 and R3 or R2 and R4 had disagreed.

For an initial (non-quantified) exploration of reasons for not using LLMs, we modeled topics with latent dirichlet allocation (LDA) using Gibbs sampling \cite{phan_learning_2008} with the R package topicmodel \cite{grun_topicmodels_2011}. We experimentally set the number of topics to 10 as a trade-off between perplexity scores and a low number of topics. We included single words and bigrams and excluded stop words, words that only occurred once, and generic words in our context, e.g., ``language model'' and ``write.'' As in \cite{eiband_when_2019}, we derived topic titles from the top 10 keywords and representative responses with high estimated proportions of topic-relevant words and bigrams.
Example responses are also from this representative set.

We report participant statements with consecutive IDs with the prefix ``N'' for non-users and ``U'' for users.

\section{Results}
First, we analyze usage and usage frequency as a function of age and gender to address hypotheses H1--H3. We then explore expertise and technology education as potential mediating factors. Finally, we categorize usage scenarios and analyze reasons for not using LLM. %

\begin{figure}%
\centering
\includegraphics[width=.5\textwidth]{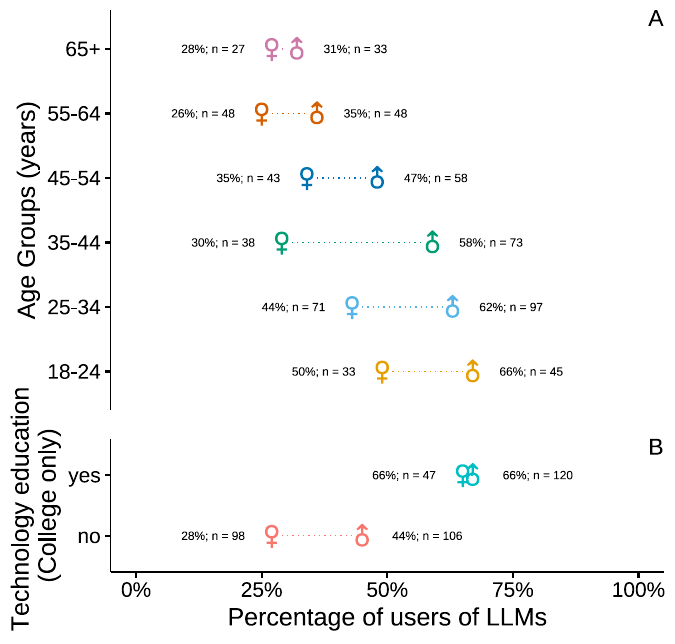}
\caption{Percentage of LLM users in a representative US sample as a function of \age{} and \gender{} (A; $ n = 1480 $) and of \textsc{Technology Education} and \gender{} for all participants with at least a college degree (B; $ n = 836 $).}
\label{fig:p1_genderageint}
\Description{Two graphs showing usage of LLMs as a function of age and gender and technology education and gender, respectively. The x-axis shows the percentage of respondents having used an LLM. The y-axis stacks the different user groups based on age and education. For each user group, the graph shows the percentage and total count of female and male respondents in that group who have used an LLM. Graph A: Age group 18-24: females 50\% (n=33) and males 66\% (n=45); 25-34: females 44\% (n=71) and males 62\% (n=97); 35-44: females 30\% (n=38) and males 58\% (n=73); 45-54: females 35\% (n=45) and males 47\% (n=58); 55-64: females 26\% (n=48) and males 35\% (n=48); 65+: females 28\% (n=27) and males 31\% (n=33). Graph B is limited to participants with a college degree. Degree not in a technology-related field: females 19\% (n=98) and males 44\% (n=120); degree in a technology-related field: females 66\% (n=47) and males 66\% (n=120). }
\end{figure}

\subsection{Age, Gender, and LLM Usage}

\paragraph{H1: \gender}
Overall, 41.5\% of participants reported to have used an LLM. We found a large gender gap with female participants being less likely to have used LLMs than male participants.  
($\female = 34.3\% < \male = 49.1 \%$; $\Delta Gender = -14.8\%$, $p_b = 0.0\%$).
Considering usage frequency for participants who had used an LLM, we again found a relatively large difference concerning gender (e.g., > once a month: $\male = 71.1\%$, $\female = 48.5\%$; $p_b = 0\%$).
Thus, the results support H1 for males and females.

\paragraph{H2: \age}
There was also a decrease in usage as a function of age group, with groups of older users being less likely to have used LLMs than younger users (18-24: 58.2\%, 25-43: 52.7\%, 35-44: 44.2\%, 45-54: 40.9\%, 55-64: 29.6\%, 65+: 29.3\%; all pairwise comparisons of posterior predictions are distinguishable at $p_b < 2.5\%$, except the comparisons of the neighboring categories 18-24 to 25-34, 35-44 to 45-54, and 55-64 to 65+). In sum, the initial usage of LLMs becomes less likely in relatively older age groups.
Interestingly, we usage in middle-aged groups using LLMs was more frequent than in young adults (55-64 > 18-24: $p_b$ = 1.95\%; 55-64 > 25-34: $p_b$ = 1.18\%; 55-64 > 25-34: $p_b$ = 1.35\%, all other $p_b$ > 2.5\%).
Thus, the results support H2 in terms of adoption but only partially for usage frequency.

\begin{figure*}%
\centering
\includegraphics[width=\linewidth]{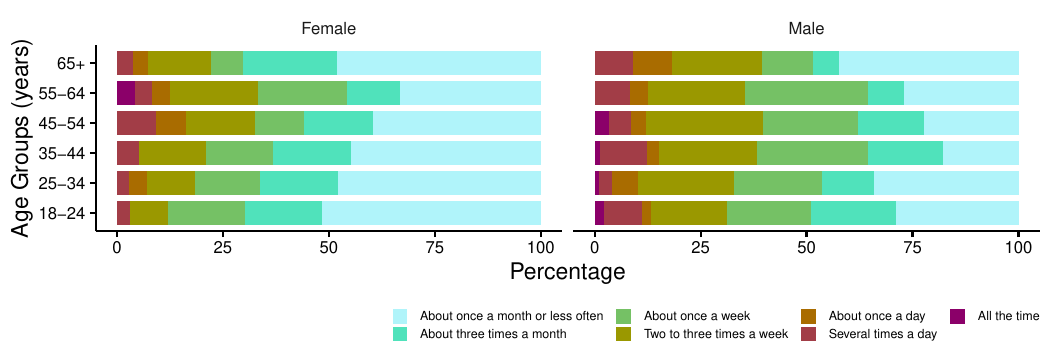}
\caption{$n = 615$; Percentage of frequency of use for LLM users in a representative US sample as a function of \age{} and \gender.}
\label{fig:p3_genderageint_freq}
\Description{Two stacked bar charts indicating the reported LLM usage frequency for female and male respondents split by age groups. Female users generally report lower usage frequencies than male users. For male users, more than 50\% across all age groups report using LLM about once a week or more often. For female users, this is only the case for the age group 55-64. For all other age groups, a majority of female users report using LLMs about three times a month or less.}
\end{figure*}

\paragraph{H3: \gender{} $ \times $ \age}
\autoref{fig:p1_genderageint}A shows the gender gap for age groups: 
While in all age groups, there is a larger share of male than female users, the discrepancy is especially large for users between 18-54 years (18-24: $\Delta Gender = -16.2 \%$, $p_b = 0.3\%$; 25-34: $\Delta Gender = -18.0\%$, $p_b = 0.0\%$; 35-44: $\Delta Gender = -27.5\%$, $p_b = 0\%$; 45-54: $\Delta Gender = -12.5\%$, $p_b = 2.0\%$) but not in old age (55-64: $\Delta Gender = -8.6 \%$, $p_b = 4.4\%$; 65+: $\Delta Gender= -2.72\%$, $p_b = 31.3\%$). %
\autoref{fig:p3_genderageint_freq} depicts the relative shares of usage frequency responses by \gender{} and \age{}.
Again, the gender gap was only present for young and middle-aged users (18-24: $p_b$ = 0.6\%; 25-34: $p_b$ = 1.2\%; 35-44: $p_b = 0.1\%$ but not for 45-54: $p_b = 7.1\% $; 55-64: $p_b = 33.8\%$; 65+: $ p_b = 8.4\% $), largely mirroring results on usage.
Thus, the results support H3: The gender gap is most salient for young and middle-aged parts of our sample, while it is not present for older participants.

\paragraph{Self-Declared Expertise}
\label{sec:expertise}

\begin{figure}%
\centering
\includegraphics[width=.5\textwidth]{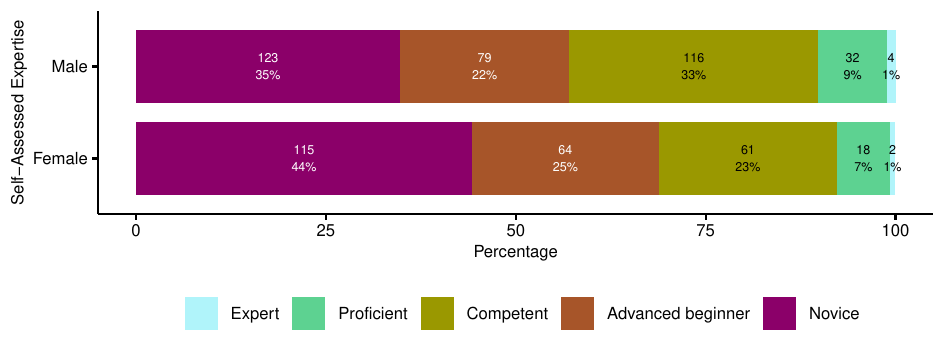}
\caption{Percentage of participants and self-declared expertise in our sample of LLM users as a function of \gender.}
\label{fig:p4_gender_llm_experience}
\Description{Two stacked bar charts indicating the self-reported level of expertise for female and male respondents, respectively. Overlays with total counts and relative frequency within the gender group. Male: Novice 123 (35\%), Advanced beginner 79 (22\%), Competent 116 (33\%), Proficient 32 (9\%), Expert 4 (1\%); Female: Novice 115 (44\%), Advanced beginner 64 (25\%), Competent 61 (23\%), Proficient 18 (7\%), Expert: 2 (1\%).}
\end{figure}

For level of expertise in using LLMs, with the five categories defined in \cite{dreyfus_peripheral_2005}, 243 participants selected ``Novice'', 181 ``Advanced Beginner'', 52 ``Competent'', 145 ``Proficient'', and only six ``Expert''.
To gain an understanding of expertise as a function of demographics, we looked at the effect of gender (the number of experts in old age was too low to warrant a more fine-grained analysis). We found that male participants rated themselves more proficient than female participants ($ p_b = 0.4\% $ in an ordinal model), see \autoref{fig:p4_gender_llm_experience}.

\paragraph{Technology Education as a Leveling Factor}
We selected participants with a college degree ($n$ = 836) and compared LLM usage depending on their discipline (technology-related or not). Adding an interaction term (\age{} $\times$ \textsc{Education}) to the statistical model, we found a similar pattern with regard to gender, age group, and their interaction as before; see \autoref{tab:models_usage}. However, the gender gap was only present for participants that did not have a technological field of studies (technological:  $\female = 66.2\% $, $\male = 66.3\% $, $\Delta Gender = -0.1\%$, $p_b =33.1\% < $ non-technological: $\female = 28.4\% $, $\male = 44.4\% $, $\Delta Gender = -15.9\% $, $p_b < 0.1\% $), see also \autoref{fig:p1_genderageint}B.

\begin{table*}%
\centering
\caption{Regression analysis. Odds with Highest-Density Interval (95\%) and $p_b$ for the usage of LLMs for two models. All variables are reference-coded (comparison female, 18-24, and no education in technology). Interaction effects are rendered as the ratios of odds ratios. Priors are on the log(odds) raw scale of the b-weights in the regression.}
\begin{tabularx}{\textwidth}{@{}p{3.8cm}ccrccrc@{}}
\toprule
\textbf{Predictors} & \multicolumn{3}{c}{\textbf{LLM usage}} & \multicolumn{3}{c}{\textbf{LLM usage (College)}} &\multicolumn{1}{c}{\textbf{Prior}} \\
\cmidrule(lr){2-4} \cmidrule(lr){5-7}
 & \textit{(Odds) ratio} & \textit{HDI (95\%)} & \textit{ $p_b$} & \textit{(Odds) ratio} & \textit{HDI (95\%)}& \textit{ $p_b$} &\textit{ (M,SD)} \\
\midrule
Intercept (female; 18-24) & 0.90 & 0.60 -- 1.32 & 29.1\% & 0.55 & 0.30 -- 0.98 & 2.1\% & $normal(0,1)$\\
Gender (male) & 2.02 & 1.21 -- 3.40 & 0.3\% & 2.12 & 1.05 -- 4.28 & 1.8\% & $normal(0,1)$\\
25-34 & 0.87 & 0.54 -- 1.42 & 29.1\% & 1.1 & 0.57 -- 2.24 & 36.9\% & $normal(0,1)$\\
35-44 & 0.50 & 0.30 -- 0.85 & 0.5\% & 0.7 & 0.34 -- 1.44 & 17.0\% & $normal(0,1)$\\
45-54 & 0.60 & 0.35 -- 1.00 & 2.4\% & 0.7 & 0.35 -- 1.53 & 20.9\% & $normal(0,1)$\\
55-64 & 0.40 & 0.24 -- 0.65 & 0.0\% & 0.5 & 0.24 -- 0.96 & 1.9\% & $normal(0,1)$\\
65+ & 0.43 & 0.25 -- 0.75 & 0.1\% & 0.71 & 0.34 -- 1.49 & 18.6\% & $normal(0,1)$\\
Gender $\times$ 25-34 & 1.03 & 0.54 -- 1.97 & 47.1\% & 1.02 & 0.44 -- 2.37 & 47.9\% & $normal(0,1)$\\
Gender $\times$ 35-44 & 1.49 & 0.75 -- 2.98 & 12.7\% & 1.30 & 0.54 -- 3.13 & 28.0\% & $normal(0,1)$\\
Gender $\times$ 45-54 & 0.83 & 0.42 -- 1.65 & 29.8\% & 1.04 & 0.43 -- 2.55 & 46.5\% & $normal(0,1)$\\
Gender $\times$ 55-64 & 0.74 & 0.38 -- 1.44 & 18.9\% & 0.71 & 0.30 -- 1.69 & 22.0\% & $normal(0,1)$\\
Gender $\times$ 65+ & 0.57 & 0.27 -- 1.29 & 7.1\% & 0.53 & 0.22 -- 1.31 & 8.4\% & $normal(0,1)$\\
Technology education & -- & -- & -- & 4.51 & 2.70 -- 7.57 & 0.0\% & $normal(0,1)$ \\
Gender $\times$ Technology education & -- & -- & -- & 0.60 & 0.32 -- 1.13 & 5.7\% & $normal(0,1)$\\
\midrule
\textit{n} & \multicolumn{3}{c}{1480} & \multicolumn{3}{c}{836} \\
\textit{R}\textsuperscript{2}  & \multicolumn{3}{c}{7.1\%} & \multicolumn{3}{c}{13.7\%} \\
\bottomrule
\end{tabularx}
\label{tab:models_usage}
\Description{Two logistic regression models predicting LLM usage of all education levels and those with a college degree, respectively. Model 1 includes the factors gender (male and female only), age, and their interaction effect.
Model 2 also includes technology and the interaction effect of technology and gender.}
\end{table*}

\subsection{Usage Scenarios}

The open-ended responses from users on typical scenarios give insights into the appropriation process, including habits that have already been established.

\paragraph{Scenario Categorization}

\begin{table*}
    \centering
    \caption{Usage scenario categorization: number of mentions per scenario category (multiple categories per response possible)}
    \small
    \begin{tabularx}{\textwidth}{>{\raggedright}X p{.3\textwidth} p{.4\textwidth} r}
    \toprule
    \textbf{Category} & \textbf{Description} & \textbf{Examples} & \textbf{Count (\%)} \\
    \midrule
    Generation in formal contexts &
        Asking the LLM to write a formal text (e.g., an email, article, letter, resume, assignment, blog entry), usage for work and school. Not generating ideas &
        \textit{Assisting me in writing emails} (U86); \textit{I refer to Chat GPT to create product descriptions for me to help my Etsy store. I also refer to it to help me word something professionally for work [...]} (U338) &
        205 (32.8\%) \\
    \midrule
    Information gain &
        searching, researching, asking questions, solving math problems, data analysis &
        \textit{I use it to help answer questions my children have about math homework.} (U441); \textit{Mostly for information on gardening and how to grow certain plants.} (U62) &
        200 (32.0\%) \\
    \midrule
    Generation in creative contexts &
        creative writing (books, poetry, fictional texts, but not jokes) and generating ideas (this can also be for formal use). Also prompt generation for other LLMs &
        \textit{[...] Creating prompts for other AI technology like Midjourney.} (U88); \textit{For inspiration in writing fiction, as well as to get ideas for cooking new meals at home. I also sometimes use it to write goofy fiction about friends, or parodies of songs [...]} (U223) & %
        141 (22.5\%) \\
    \midrule
    Trying it out &
        Trying an LLM out, ``only used it once'', experimenting with prompts, beta testing an LLM &
        \textit{i honestly just wanted to try it out because you hear so much about them in the news} (U24); \textit{To ask questions because I'm curious about how it operates} (U327) &
        75 (12\%) \\
    \midrule
    Entertainment &
        Writing jokes, ``just to play around with it'' &
        \textit{For entertainment (e.g. ``write me a song'') with friends}; \textit{It helped me generate scenarios for a roleplaying game.} (U274); \textit{for entertainment. I will ask pointless questions to explore responses.} (U242) &
        56 (8.9\%) \\
    \midrule
    Coding &
        Coding and debugging, asking for programming support, training an AI model &
        \textit{Getting help writing scripts and other code. [...]} (U454); \textit{Designing warehouse and IT tasks for AWS.} (U54) & %
        41 (6.5\%) \\
    \midrule
    Summarization \& Explanation &
        Summarizing, paraphrasing, simplifying, or asking for explanations of a topic. Not information gain or instructions &
        \textit{I will typically use them to summarize clinical studies and scientific studies.} (U34); \textit{I use them to explain concepts when I want to compare different ideas that I am not an expert on.} (U26) &
        37 (5.9\%) \\
    \midrule
    Recommendation &
        Suggestions or recommendations for travel destinations, movies, actions to take, etc. Not requesting ideas &
        \textit{Create playlists, theoretical hockey teams, beer recipes} (U119); \textit{Ask them to give me a workout tip.} (U570) &
        34 (5.4\%) \\
    \midrule
    Messaging \& Social Media &
        Informal social media posts or messaging content. Not emails or blogs &
        \textit{To create content for social media.} (U280); \textit{Customer service} (U331) &
        33 (5.3\%) \\
    \midrule
    Chatting with AI &
        Discussing or interacting with a chatbot &
        \textit{chatbots on websites} (U233); \textit{Get ``moral support'' by expressing my issues and getting a response even if it's not from a human.} (U100) &
        22 (3.5\%) \\
    \midrule
    Editorial &
        Proofreading, suggesting edits to a text (excl. code), rewriting something &
        \textit{[...] summarize or expand texts, and adapt the style or tone of a text.} (U424); \textit{for my homework assignments to rewrite my answers better} (U518) &
        23 (3.7\%) \\
    \midrule
    Language &
        Translating or transcribing text, learning a language &
        \textit{I'm learning to speak Spanish and I use it to help me translate Spanish to English when I'm stumped.} (U422) &
        21 (3.4\%) \\
    \midrule
    Studies (tasked) &
        Participation in a study where the use of LLM was a part of the study &
        \textit{Working on surveys with AI prompts on Prolific} (U372) &
        11 (1.8\%) \\
    \midrule
    Studies (unclear) &
        Using an LLM to answer questions in a study (not tasked or ambiguous regarding if tasked) &
        \textit{Answer an open ended basic question on a survey, like, how do you feel today?} (U377) &
        7 (1.1\%) \\
    \midrule
    Ambiguous &
        unclear, ambiguous, or incomprehensible scenarios, indicating they haven't used an LLM (despite survey branching), suspected LLM-generated responses &
        \textit{I don’t use them} (U391); \textit{Reading Facebook post} (U433); \textit{In completing tasks} (U490) &
        40 (6.4\%) \\
    \bottomrule
    \end{tabularx}
    \label{tab:usage_scenarios}
\end{table*}

\autoref{tab:usage_scenarios} summarizes the inductively derived usage scenario categorization and example responses from LLM users to the question \textit{What are typical tasks or scenarios where you use large language models?}.
The examples span a wide range from professional applications to entertainment. The most prevalent scenarios were text generation in formal contexts and information gain. At the current state of adoption, a number of respondents were also simply testing the capabilities of LLMs.
Some participants mentioned very precise use cases, e.g., summarizing clinical studies (U34) or writing product descriptions for an Etsy store (U338).
Interestingly, several participants had used LLMs in the context of user studies. In some cases, it was clear that LLMs were a part of a survey. However, responses like the one given by U377 indicate that LLMs are also employed to automate study participation.

\paragraph{Differences for Novices and Proficient/Expert Users}
As a second step, we performed an exploratory regression analysis predicting the occurrence of selected scenario categories from expertise levels (cf. \autoref{sec:expertise}). Notably, we expected that experts would be more likely to use LLMs in professional contexts, while non-targeted use would be more likely for novices.
Because there were only six experts and 51 proficient users, we merged the top three self-declared expertise levels into \textit{Competetent, Proficient \& Expert} ($n = 238 $) for comparing scenario categories to those mentioned by \textit{Novice} users ($ n = 243$).
\autoref{tab:expertise_scenario_models} shows that proficient users are more likely to mention scenarios categorized as generation of formal texts and coding, while novice users are more likely to try them out. Entertainment is only descriptively slightly more frequent for novices.
The second-most mentioned category, information gain, is equally prominent for both novices and proficient users.

\begin{table*}
    \centering
    \caption{Relative frequency and frequency difference of responses from the different usage scenario categories and analysis of Bayesian logistic regression models, for novice vs. competent, proficient, and expert users. Based on paired differences of 4000 draws from the posterior distributions, we report the Highest-Density Interval (95\%), the respective odds ratio, and $ p_b $.}
    \begin{tabular}{l c c c c c c c}
        \toprule
        \textbf{Category} & \textbf{Novice} & \textbf{Competent -- Expert} & $ \Delta $ & \textbf{HDI (95\%)} & \textbf{Odds ratio} & $ p_b $ & \textbf{Prior}\\
        \midrule
        Information gain & 32.5\% & 31.9\% & -0.6\% & -9\% -- 8\% & 1.02 & 44.6\% & $normal(0,1)$ \\
        Generation formal & 25.1\% & 36.6\% & 11.5\% & 3\% -- 19\% & 0.69 & 0.4\% & $normal(0,1)$ \\
        Trying it out & 18.1\% & 8.4\% & -9.9\% & -15\% -- -3\% & 2.05 & 0.1\% & $normal(0,1)$ \\
        Entertainment & 11.9\% & 6.7\% & -5.2\% & -10\% -- 0\% & 1.70 & 2.9\% & $normal(0,1)$ \\
        Coding & 1.7\% & 10.9\% & 9.2\% & 4\% -- 12\% & 0.23 & 0.0\% & $normal(0,1)$  \\
        \bottomrule
    \end{tabular}
    \label{tab:expertise_scenario_models}
\end{table*}

\subsection{Reasons for Not Using LLMs}

\autoref{tab:lda_topics} summarizes reasons for not using LLMs as derived with LDA topic modeling, revealing barriers to adoption. These include a general lack of knowledge of LLMs and their usage, a preference for human writing and creativity, and concerns regarding plagiarism, cheating, and accuracy.
Thus, non-users consider both individual and societal aspects.

\begin{table*}
    \centering
    \caption{LDA topic modeling of open-ended responses from non-users for the question \textit{What are your reasons for not using large language models?} We merged Topics 2 and 4 because the examples were very similar.}
    \small
    \begin{tabularx}{\textwidth}{>{\raggedright\arraybackslash}X >{\raggedright\arraybackslash}p{.32\textwidth} p{.5\textwidth}}
        \toprule
        \textbf{Topic} & \textbf{Representative tokens} & \textbf{Examples} \\
        \midrule
        No need &
            ai, learn, people, world, busy, issues, daily life, aware, plagiarism, application, ideas &
            \textit{Normal everyday life does not call for these.} (N554); \textit{I am a writer by training, so I don't need to use an AI for that. I'm sure I'm using AI daily in other ways, but LLM aren't something I have seen a need for in my daily life at this point.} (N732) \\
        \midrule
        Lack of knowledge &
            Topic 2: heard, school, opportunity, moment, knowledge, words, reasons, stereotypes, research, easier, read, idea, mind, difficult \newline Topic 4: lack, prefer, knowledge, concerns, job, cheating, ethical, english, level, aware, learn, computer, voice, idea, term &
            \textit{No idea of them prior to this} (N765); \textit{My lack of knowledge on how to use one as well as my lack of need for one at the moment.} (N863); \textit{Just never had the opportunity} (N403); \textit{I don't fully understand it's purpose or the ways in which it could benefit me. Lack of knowledge.} (N721) \\
        \midrule
        Prefer own/human writing &
            information, ai, daily, current, humans, retired, guess, accurate, errors, chat gpt, google &
            \textit{I am a good writer and am pretty efficient at putting my thoughts into words} (N288); \textit{The writing would not be in my style or display my personality.} (N714); \textit{I don't like the idea of people depending on such technology. I'm concerned for the power AI could have one day.} (N80) \\
        \midrule
        Human touch &
            life, human, existed, data, information, experience, writer, benefit, privacy concerns, machines, authentic, paper, everyday life, bit &
            \textit{Because it’s not authentic human speech(writing)} (N238); \textit{Too impersonal. I would prefer to talk to an actual person} (N419) \\
        \midrule
        Plagiarism \& trust &
            time, access, papers, heard, sound, gpt, simply, training, technology, trust, existed, school, essays, pretty, class, writing papers, create, terms, unethical &
            \textit{If I was to do a paper for a class in school, I wouldn't want to use it because it wouldn't be my own work.} (N335); \textit{I don't trust how they source things.} (N102) \\
        \midrule
        Intellect &
            understand, cheating, person, chance, bias, writing skills, original, prefer, stuff, words, impact, chat, text, student &
            \textit{I think that it dumbs down people that could otherwise learn and improve their own writing skills.} (N605); \textit{I can write original stuff, unlike people who use ChatGPT.} (N2) \\
        \midrule
        Cheating \& lack of creativity &
            feel, familiar, personal, creativity, creative, generated, social, coding, replace, machine, biased, havent, required &
            \textit{It seem like something a person should do themselves. Using the algorithm feel like cheating.} (N379); \textit{Limiting the need for human creativity.  Taking away initiative} (N204); \textit{It reduces human creativity} (N520) \\
        \midrule
        Privacy \& accuracy concerns &
            trust, privacy, people, accurate, students, ability, cost, harmful, personally, content &
            \textit{Lack of trust and invasion of privacy} (N588); \textit{I don’t trust them to actually be accurate and they aren’t able to accurately convey the human's emotion} (N88); \textit{I might not use them when I needed accurate answers.  As of now, these models are not capable of always providing the correct answers to prompts.} (N317) \\
        \midrule
        Values \& abuse &
            technology, real, familiar, time, feels, creativity, opportunity, learning, written, knowing &
            \textit{It'll be abused. For example, there are foreign companies that make money by scamming people (eg romance scams). [...] Also, it feels cold and soulless as it lacks a real person behind it. [...]} (N397); \textit{I think it is a scary technology that could be a problem for our democracy. We will not be able to distinguish real/fake or true/untrue.} (N584) \\
        \bottomrule
    \end{tabularx}
    \label{tab:lda_topics}
\Description{Overview of latent dirichlet allocation on reasons for not using LLMs. Columns: Themes, top 10 keywords (more in case of ties) and exemplary participant statements.}
\end{table*}

\section{Discussion}

Equitable technological transformation and accessibility are paramount challenges in the digital transformation of society: specific demographic profiles are left out, and this raises the question of digital inequality \cite{robinson2015}.
Our analysis guides researchers in detecting and mitigating inequalities in LLM adoption and in designing for diverse usage scenarios.

\subsection{Promoting Equity in LLM Adoption}
In a representative sample of the US population, we found a substantial gender gap (H1) and age effect (H2) for LLM use. Female gender and old age were related to a decreased likelihood of having used LLMs. 
Our findings align with similar trends of other disruptive digital technologies such as the world wide web \cite{robinson2015} %
or immersive technologies \cite{felnhofer2012virtual}. In line with Robinson et al. \citep{robinson2015}, the socio-demographic disparities of LLM use reflect social digital inequalities in an amplified manner. 

Nonetheless, our research diverges from previous studies on the demographics of technology use. Namely, the gender gap was most pronounced in young and middle-aged adults and was non-existent in old age. This contradicts simple age effects where young age correlates with being tech-savvy and quickly adopting new technologies. By investigating self-rated expertise in our sample, we find that female participants report lower expertise levels, which potentially further delays their adoption of LLMs. Thus, our more granular analysis of the demographic profiles revealed interaction patterns of age and gender (H3). This presents a unique challenge to address the gender gap as early as possible. 

Importantly, we found no differences in LLM use between male and female respondents who had received technology education in college. As AI systems become more integrated into working life, there is a risk that gender disparities in AI skills could increase inequality \cite{AIWorkingLivesofWomen}. Our results underscore the importance of education and suggest that education in technology is crucial in bridging the digital gender gap for facilitating equitable digitization of society.
This conclusion is in line with recommendations forwarded by international organizations, such as IDB, OECD, and UNESCO \cite{AIWorkingLivesofWomen, west2019d}.

\subsection{Mitigating Barriers to LLM Adoption}

The reasons for not using LLMs that participants mentioned reveal several concrete barriers to widespread adoption beyond quantified demographic factors. Tackling them requires improvement in LLM algorithms and interfaces as well as usage education. Below, we list possible ideas for addressing key barriers.

\paragraph{Respecting Ethics and Values}
Several non-users felt that using LLMs is cheating and promotes plagiarism.
This attitude is likely reinforced by critical news articles\footnote{e.g., \url{https://www.cnbc.com/2023/06/06/news-organizations-ai-disinformation.html}} and the guidelines on LLM usage that educational institutions and professional organizations around the globe are currently putting forward (e.g., the ACM's policy on authorship\footnote{\url{https://www.acm.org/publications/policies/new-acm-policy-on-authorship}}).
However, the ethical impact depends substantially on the concrete usage context: presenting an LLM-generated essay as one's own is clearly more problematic than asking an LLM to generate scenarios for a role-playing game or requesting information on gardening (cf. \autoref{tab:usage_scenarios}). Even within educational contexts, LLMs have a lot of potential, e.g., for generating educational material and fostering student motivation \cite{kasneci_chatgpt_2023} or as personalized tutors \cite{mollick_assigning_2023}.
Therefore, effective communication on LLMs should clearly delineate different applications contexts and tasks of LLMs to allow a differentiated ethical assessment.

\paragraph{Allowing for Verification of Accuracy}
Hallucinations remain a challenge in natural language generation \cite{ji_survey_2023} that contributes to an impression of inaccuracy. Current work is exploring mitigation options in both the front-end and back-end. For example, interface layers could enable fact-checking \cite{leiser_chatgpt_2023}, and integrating expert tools into LLM-based systems could improve the output accuracy \cite{bran_chemcrow_2023}.

\paragraph{Protecting Privacy}
Privacy risks were another concern. While the actual privacy risk depends on the LLM instance, there have been examples of data breaches\footnote{e.g., \url{https://openai.com/blog/march-20-chatgpt-outage}}. If not deactivated, ChatGPT may use recorded conversations for re-training the system\footnote{\url{https://help.openai.com/en/articles/7730893-data-controls-faq}}, which proved critical for Samsung when employees shared confidential data in their chats\footnote{\url{https://mashable.com/article/samsung-chatgpt-leak-details}}.
Training opt-out options should be made more visible to improve users' perception and control of privacy. Running isolated model instances may also be advisable in high-risk domains.

\paragraph{Catering to the Need for Autonomy}
The responses revolving around preferring the participants' own writing indicate a need for autonomy in writing and conversely, a feeling of being overruled by an AI.
This is closely related to car drivers not willing to rely on assistance systems \cite{reagan_crash_2018} %
or computer users sticking to basic editors such as vim and emacs\footnote{\url{https://stackoverflow.blog/2020/11/09/modern-ide-vs-vim-emacs/}}, valuing principles and habits over assistance.
To satisfy the need for autonomy, we suggest clarifying that using LLMs does not mean giving up all writing tasks. A good option may be to instruct LLMs to operate as tutors that do not directly provide a solution to a problem, but rather ask questions or provide hints that help one discover the solution oneself \cite{khan2023how}. Again, context dependence is key and should reflect individual needs and priorities, e.g., using LLMs primarily for non-personal tasks.

\paragraph{Promoting Usage Education}
Participants who feel unsure about using LLMs will need usage education to become sufficiently confident. Previous work on technology adoption has shown that support from interpersonal relationships can accelerate adoption \cite{magsamenconrad_mobile_2020}. Therefore, establishing support communities could facilitate LLM usage.

\paragraph{Promoting the Opportunities of Using LLMs}
People who currently do not see any benefit in using LLMs could be provided with showcases of LLM results in diverse scenarios (cf. \autoref{sec:supporting_usage}). Besides guidance and templates, interpersonal relationships could also facilitate adoption, e.g., younger users supporting older users \cite{magsamenconrad_mobile_2020}.

\subsection{Supporting Effective LLM Usage}
\label{sec:supporting_usage}

The usage categorization revealed formal generation, information gain, and creative generation as prevalent.
Note that there may be overlaps or hierarchical dependencies in the categories we derived. For example, coding could also be considered generation in a formal context. However, we kept it as a standalone category because it has received a lot of attention in LLM training (e.g., \cite{wu_promptchainer_2022}) and is one of the first domains for specialized interfaces such as Copilot.

Some of the scenarios were already quite specific and creative, e.g., building theoretical hockey teams (U119). This indicates that more experienced respondents have already managed to appropriate the technology to match their individual needs \cite{salovaara_everyday_2011}.
In contrast, many respondents also reported that they primarily try out LLMs. As usage patterns stabilize, we expect this share to decrease.

Contrasting usage patterns by expertise showed that proficient and expert users were more likely than novice users to use LLMs for professional tasks, including generation in formal contexts and coding.
This implies that novice users may be disadvantaged in the labor market; as \citet{eloundou_gpts_2023} remark, policymakers should consider the societal impact of large language models.
The differences in usage contexts could also inform support strategies for less experienced users. For example, scenario descriptions and example prompts, as recently implemented in ChatGPT, give users an idea of the potential of LLMs.

\subsection{Future of User Research}
The rise of LLMs also has implications for conducting online studies like ours. Our data corroborates recent predictions and emerging evidence that textual research data collected online may be partially generated by LLMs \cite{hamalainen2023evaluating,veselovsky2023artificial}. We had to exclude two participants with clearly LLM-generated responses such as ``Large language models like me can be used to [...].'' Further, 7 of our participants reported they have used LLMs to answer questions in a study, using wordings that suggested LLM use had not been instructed. This trend towards automated responses means that we need to rethink the role and design of online surveys in user research. %
Because fully reliable detection of LLM-generated text does not seem possible \cite{weber2023testing,sankar2023can,lu2023large}, user researchers may need to reconsider other, more traditional data collection approaches such as lab studies or (automated) telephone surveys. Alternatively, online participants should be monitored to ensure that they are not using LLMs, which is problematic both technologically and from a privacy perspective.

\subsection{Limitations}

Our sample showed an absolute level of usage of 41.5\%, suggesting that an early majority of the population has already used LLMs. This number might be confounded by the online context of the study and the fact that we did not differentiate by LLM instances.
It will be subject to change as LLMs are incorporated into the software products we use on a daily basis. 
Nevertheless, we expect that the detected socio-demographic disparities are not specific to our study setting and may even be amplified for offline communities.
Still, as LLMs expand to societies globally, data from other countries is needed for making robust global estimates of usage. For example, we expect the representation of different value systems in LLMs to influence adoption \cite{johnson_ghost_2022}. %
Future work should also monitor the LLM adoption process with regard to demographic and educational factors to evaluate the need for targeted interventions as early as possible.

\section{Conclusion}

Large language models have disrupted the technology market within a few months, enabling novel kinds of AI support in a variety of professional and non-professional contexts alike. The simple chat-style interfaces promise equitable access. However, young males are currently the predominant user group of LLMs in our representative US sample. This shows that already existing imbalances in a technology-driven society are reinforced. Moreover, users with higher levels of expertise are more likely to use LLMs in professional contexts, which may further increase their efficiency -- that less technology-affine people cannot compete with. Non-users are held back by issues such as ethical concerns and conflicting values, inaccurate LLM output, but also a lack of knowledge and opportunities.
Nonetheless, our findings also show that the gender gap is less pronounced in higher age and that technology-related education levels the differences in LLM adoption.
Therefore, we call for removing barriers to LLM adoption. Important steps are supporting technology education, promoting ethical and privacy-preserving use, addressing inaccuracies and biases in LLMs, and targeting support structures toward disadvantaged demographic profiles.
Future studies should expand our analyses beyond the US to cater to specific needs across cultural backgrounds.

\begin{acks}
Will be inserted later.
\end{acks}

\bibliographystyle{ACM-Reference-Format}
\bibliography{sample-base}

\end{document}